\documentclass[preprint,showpacs,preprintnumbers,amsmath,amssymb]{revtex4}
\usepackage{graphicx}
\usepackage{dcolumn}
\usepackage{bm}
\usepackage{epsfig}
\def\trade{{\bigcirc}\!\!\!\!\!\mbox{{\tiny R}}}
\def\mathmath{{\it Mathematica}$_{\trade}$\,}
\newcommand{\geqnew}{\stackrel{>}{\!\ _{\sim}}}
\newcommand{\leqnew}{\stackrel{<}{\!\ _{\sim}}}
\begin{document}

\preprint{PSU/TH-252}

\title{Wave packet revivals and the energy eigenvalue spectrum 
\\ of the quantum pendulum}

\author{M. A. Doncheski} \email{mad10@psu.edu}
\affiliation{%
Department of Physics\\
The Pennsylvania State University \\
Mont Alto, PA 17237 USA \\
}%

\author{R. W. Robinett} \email{rick@phys.psu.edu}
\affiliation{%
Department of Physics\\
The Pennsylvania State University\\
University Park, PA 16802 USA \\
}%

\date{\today}

\begin{abstract}

The rigid pendulum, both as a classical and as a quantum problem,
is an interesting system as it has the exactly soluble harmonic oscillator 
and the rigid rotor systems  as limiting cases in the low- and high-energy 
limits respectively. The energy variation of the classical periodicity 
($\tau$) is also dramatic, having the special limiting case of 
$\tau \rightarrow \infty$ at the 'top' of the classical motion 
(i.e. the separatrix.)
We study the time-dependence of the quantum pendulum problem, focusing on 
the behavior of both the (approximate) classical periodicity and especially
the quantum revival and superrevival times, as encoded in the energy
eigenvalue spectrum of the system. 
We provide approximate expressions for the energy eigenvalues in both
the small and large quantum number limits, up to 4th order in perturbation
theory, comparing these to existing handbook expansions for the 
characteristic values of the related Mathieu equation, obtained by other
methods. 
We then use these approximations to probe the classical periodicity, as 
well as to extract information on the quantum revival and 
superrevival times. We
find that while both the classical and quantum periodicities increase 
monotonically as one approaches the 'top' in energy, from either above or 
below, the revival times decrease from their low- and high-energy values 
until very near the separatrix where they increase to a large, but finite 
value.
\end{abstract}

\pacs{03.65.Ge, 03.65.Sq}

%

\maketitle

\section{\label{sec:intro} Introduction}

The problem of a point particle, restricted to a circular radius, and
subject to a uniform gravitational (or electric) field, namely a rigid 
pendulum, is one of the most familiar of 
all problems in classical mechanics. 
The corresponding quantum version was first considered 
early in the history
of quantum mechanics by Condon \cite{pendulum_condon} and is sometimes
discussed in the more physical context of hindered internal rotations
\cite{hindered_book1} - \cite{pendulum_baker} in chemistry. The problem
is also a staple of the pedagogical literature \cite{pendulum_khare}
- \cite{pendulum_aldrovandi} and is often seen as an example of
perturbation theory methods \cite{pendulum_martin} 
- \cite{pendulum_kiang},
especially in collections of problems in quantum mechanics
\cite{problem_const} - \cite{problem_cronin}, where is often described
as the plane rotator in an electric field. Most treatments focus 
attention on aspects of the structure of the energy eigenvalues spectrum,
with only a handful (see, {\it e.g.}, \cite{pendulum_cook}) examining the
question of time-dependence of wave packet solutions.

More recently, the realization that important information on longer-term
quantum correlations, in the form of wave packet revivals and 
superrevivals,  is also encoded in the energy eigenvalue spectrum, 
has focused new attention on the time dependence of physical
and model systems which exhibit such interesting dynamics. In many such
systems, initially localized states which have a short-term, 
quasi-classical 
time evolution, can spread significantly over several orbits, 
only to reform 
later in the form of a quantum revival in which the spreading reverses 
itself, the wave packet relocalizes, and the semi-classical periodicity is 
once again evident. Such revival phenomena have been observed in a wide 
variety of physical systems, especially in Rydberg atoms 
\cite{revival_book}, \cite{revival_review},  and calculations exist for 
many other model systems \cite{other_revivals}.

The  archetype of a one-dimensional model system for quantum  revivals
is the infinite well (where such revivals are exact) and a number
of analyses \cite{1d_fractional} - \cite{styer}
have provided insight into both the short-term and long-term 
behavior of wave packets. The rigid rotor, which like the infinite
well, also has an $n^2$ quantum number dependence in the energy spectrum,
behaves very similarly \cite{bluhm}, \cite{canadian}, 
with neither system exhibiting higher-order
correlations (such as superrevivals) due to the simplicity of their
energy level structure. Harmonic oscillator systems, with a simple
spectrum proportional to $n$, do not exhibit quantum revivals (or rather
have $T_{rev} \rightarrow \infty$), or super-revivals, 
but Morse-type generalizations
\cite{morse}, \cite{carpets}, with anharmonic terms present,  
have predictable revival structures, as confirmed in 
observations of molecular systems \cite{molecular}. One-dimensional
power-law potentials of the form $V_{(k)}(x) = V_0|x/L|^{k}$ also provide
examples \cite{robinett_jmp} of interesting quantum revival time behavior 
as a model parameter is varied, as they interpolate
between the cases of the oscillator ($k=2$) and infinite well 
($k \rightarrow \infty$) limits.

For many such systems, which depend on a single quantum
number, one typically expands the energy eigenvalues (assuming integral
values) about the central value used in the construction of a wave packet
via 
\begin{equation}
E(n) \approx E(n_0) + E'(n_0)(n-n_0) + \frac{1}{2}E''(n_0)(n-n_0)^2
+ \frac{1}{6}E'''(n_0)(n-n_0)^3 + \cdots
\end{equation}
in terms of which the classical period, revival, and superrevival times
are given respectively by
\begin{equation}
\tau \equiv T_{cl} = \frac{2\pi\hbar}{|E'(n_0)|}
\quad 
,
\qquad
T_{rev} = \frac{2\pi \hbar}{|E''(n_0)|/2}
\quad
, 
\qquad 
T_{super} = \frac{2\pi \hbar}{|E'''(n_0)|/6}
\label{period_definitions}
\end{equation}
so that information on the long-term quantum correlations present in many
wave packet solutions is encoded in the energy spectrum, just as is the 
short-term quasi-classical time-dependence. In typical experimental
realizations of such systems, one finds $\tau << T_{rev} << T_{super}$.

In this report, we will focus on the structure of the quantum revival
 (and superrevival) time for the problem of the quantum pendulum,
extending earlier analyses of the approximate energy spectrum for this
problem, in both the low- and high-energy limits, as well as making use
of numerically obtained values for the entire spectrum. We are interested
in such questions as (i) what is the revival time, $T_{rev}$ for the 
(anharmonic) oscillator limit of the quantum pendulum and (ii), 
how do the quantum revival and superrevival times vary (from both the 
low-energy oscillator and high-energy rotor limits) as one approaches 
energies near the separatrix value. To probe such questions, we begin,
in Sec.~\ref{sec:classical_background} by defining the generic problem of 
the quantum pendulum, and review standard expressions for the energy 
dependence
of the classical periodicity of the pendulum, $\tau(E)$ versus $E$.
We then discuss, in Sec.~\ref{sec:mathieu}, the quantum version of this
problem, reviewing its equivalence to the 
Mathieu equation of the mathematical
physics literature, and making contact with 'handbook' results 
for its periodic solutions from standard references. 
In Secs.~\ref{sec:high_energy}
and \ref{sec:low_energy}, we discuss perturbative solutions to the quantum 
pendulum problem, in the high energy (free rotor) and low energy (oscillator
like) limits respectively, deriving results up to 4th order in 
perturbation theory.
We use these results to discuss the classical periodicity, as well as
the quantum revivals and superrevivals in these two limiting cases.
Finally, we use these earlier results as benchmarks against which to
compare the numerical evaluation of the energy eigenvalues for the
pendulum over the entire spectrum, focusing on the behavior of
$T_{rev}$ and $T_{super}$ near the 'top' of the pendulum. 
Some of the mathematical details of the high-energy limit are discussed
in a WKB-approximation in Appendix~\ref{sec:wkb_appendix}, while the
details of the perturbative expansion in the low-energy, oscillator-like
limit are given in Appendix~\ref{sec:third_and_4th_appendix}.

\section{\label{sec:classical_background} Classical background}

For definiteness, we will consider a point particle of mass $\mu$
(a notation used to avoid confusion with familiar angular momentum 
quantum numbers) restricted to a circle of radius $l$, which therefore
has a rotational inertia $I \equiv \mu l^2$. It is subject to either a
uniform gravitational ($mg$) or electric ($QE$) force in the vertical 
direction, giving rise to a potential energy function of the form
\begin{equation}
V(\theta) = -V_0\cos(\theta)
\qquad
\mbox{where}
\qquad
     V_0  = \left\{ \begin{array}{ll}
               mgl & \mbox{(quantum pendulum)} \\
               QEl & \mbox{(Stark effect for rigid rotor)}
                                \end{array}
\right.
\, . 
\end{equation}
(One can also consider a point electric dipole, $p$, in a uniform 
electric field, in which case $V_0 = p E$.)

We can rewrite the classical conservation of energy connection
\begin{equation}
\frac{I}{2}\left(\frac{d\theta}{dt}\right)^2 - V_0\cos(\theta) = E
\end{equation}
in the form
\begin{equation}
\frac{d\theta}{dt} = \sqrt{\frac{2}{I}} \sqrt{E + V_0\cos(\theta)}
\end{equation}
or
\begin{equation}
\sqrt{\frac{I}{2}} \frac{d\theta}{\sqrt{E+V_0\cos(\theta)}} = dt
\, 
\end{equation}
which can, of course,  be formally integrated in terms of elliptic 
integrals (or evaluated numerically) to obtain $\theta(t)$ solutions.

For energies satisfying $-V_0< E< +V_0$, the motion will be periodic
with classical turning points at $\pm \Theta = \pm \cos^{-1}(-E/V_0)$
and the classical period is given by
\begin{equation}
\tau(E)  = 2 \sqrt{\frac{I}{2}} \int_{-\Theta}^{+\Theta}\,
\frac{d\theta}{\sqrt{E+V_0\cos(\theta)}}
\label{low_energy_period}
\end{equation}
where the extra factor of $2$ comes from including both the 'back' and the
'forth' motions. 
For low energies,  in the limit of small oscillations,  
with $-V_0 \leqnew E << +V_0$ 
or $\Theta << 1$,  the motion is oscillator like 
with an energy independent 
period given by $\tau = 2\pi/\sqrt{V_0/I}$. In the $E \rightarrow +V_0$
(or separatrix) limit, the classical period diverges logarithmically. 

For the case
of unbounded motion, where $E>+V_0$, one period corresponds more simply
to a single revolution, with the result
\begin{equation}
\tau(E) = \sqrt{\frac{I}{2}} \int_{-\pi}^{+\pi} 
\frac{d\theta}{\sqrt{E+V_0\cos(\theta)}}
\label{high_energy_period}
\end{equation}
and in the high-energy (rotor-like) limit,
when $E\equiv I\Omega^2/2 >> +V_0$, 
we find that
\begin{equation}
\tau \rightarrow \sqrt{\frac{I}{2}}\frac{2\pi}{\sqrt{E}} = 
\frac{2\pi}{\Omega}
\label{classical_limit}
\end{equation}
 as expected.
The expressions in Eqn.~(\ref{low_energy_period}) and
(\ref{high_energy_period}) for $\tau(E)$ versus $E$ 
can then be compared to the corresponding quantum values of
$\tau_{n}(E_n)$ versus $E_{n}$ in the appropriate limiting cases.

\section{\label{sec:mathieu} Schr\"odinger equation for 
the quantum pendulum and Mathieu equation}

The Schr\"odinger equation for the quantum pendulum can be written in terms
of the moment of inertia, $I$, in the form
\begin{equation}
\frac{\hat{p}_{\theta}^2}{2I}\psi(\theta) + V(\theta)
\psi(\theta) =
-\frac{\hbar^2}{2I} \frac{d^2\psi(\theta)}{d\theta^2}
- V_0 \cos(\theta) \psi(\theta)
= E \psi(\theta)
\, .
\label{basic_equation}
\end{equation}
The conventional change of variables, $\theta \equiv 2z$,  can be 
used to rewrite this as
\begin{equation}
\frac{d^2 \psi(z)}{dz^2} + \left[ \frac{8IE}{\hbar^2}
+ \frac{8IV_0}{\hbar^2}\cos(2z)\right]\psi(z)
\end{equation}
which is recognizable as a familiar equation of mathematical physics,
\begin{equation}
\frac{d^2\psi(z)}{dz^2} + [a + 2q\cos(z)] \psi(z)
\end{equation}
namely Mathieu's equation \cite{mac_mathieu}, \cite{stegun}
(with $q \rightarrow -q$ in this case) and where we identify
\begin{equation}
q \equiv \frac{4IV_0}{\hbar^2}
\qquad
\mbox{and}
\qquad
a \equiv \frac{8IE}{\hbar^2}
\, .
\label{mathieu_parameters}
\end{equation}
Because of the intrinsic parity of the potential, the solutions can
be characterized as being even, $ce_{r}(z,q)$ or cosine-like, or
odd, $se_{r}(z,q)$ or sine-like, for integral values of $r$, with 
$r\geq 0$ ($r\geq 1$) for even (odd) cases. 
For the limiting case of $q=0$, the solutions 
(in standard normalization \cite{mac_mathieu}, \cite{stegun})
have the forms
\begin{eqnarray}
ce_{0}(z,0) & = & 1/\sqrt{2} \qquad r = 0 \\
ce_{r}(z,0) & = & \cos(rz)   \qquad r \geq 1\\
se_{r}(z,0) & = & \sin(rz)   \qquad r \geq 1
\, . 
\end{eqnarray}

For a fixed value of $q$, only for specific (or in more standard notation, 
characteristic) values of the parameter 
$a$ will the solutions be periodic, with periods of $\pi$ or $2\pi$
in the variable $z$, and these are denoted by $a_{r}, b_{r}$ for the even 
and odd solutions 
respectively. For each value of $q$, there are countably infinite number
of solutions, labeled by $r$. For the physical case considered here,
where the system must exhibit $2 \pi$ periodicity in the  variable
$\theta$, one chooses the even integer ($r=2m$) $ce_{2m}(z,q)$ and 
$se_{2m}(z,q)$ solutions which have period $\pi$ in 
the variable $z$. Standard mathematical packages (such as \mathmath)
can quickly evaluate the characteristic values for a given value of $q$ and
we plot in Fig.~1 the pattern of characteristic values $a_{2m}$ 
(solid curves for the even solutions) and $b_{2m}$ (dashed curves for 
the odd solutions)
which are proportional to the allowed quantized energies, $E$, for 
different values of $q$, corresponding to increasing values of $V_0$. 
For $E>> +V_0$ ($a>>q$), the solutions are doubly degenerate corresponding
to the free rotor limit: in this case, the real $ce_{2m}$ and $se_{2m}$ 
solutions are analogous to linear combinations of 
clockwise ($\exp(im\theta)$)
and counterclockwise ($\exp(-im\theta)$) rotational eigenstates of the 
plane rotator.
For $-V_0 < E << +V_0$, the spectrum splits into equally spaced oscillator
like states corresponding to the quantized version of small angle
isochronous oscillations.

Standard mathematical references \cite{mac_mathieu}, \cite{stegun} provide
approximations for the characteristic values, $a_{r},b_{r}$, as a function
of $q$ in both the $q<<1$ and $q>>1$ limits, typically derived by
expanding the solutions in appropriate Fourier-like series expansions. For
the 'high energy' or rotor limit, with $q<<1$, one finds 
(for $r \geqnew 7$)
that the $a_{r},b_{r}$ are approximately degenerate with
\begin{equation}
\left.
\begin{array}{l}
a_{r} \\
b_{r} \end{array}
\right\}
= r^2 + \frac{q^2}{2(r^2-1)} + \frac{(5r^2+7)q^4}{32(r^2-1)^3(r^2-4)}
+ \cdots
\,\,\, .
\label{mathieu_limit_high}
\end{equation}
and the difference between the characteristic values for 
even and odd solutions is known to satisfy 
\begin{equation}
a_{r}-b_{r} = {\cal O}(q^r/r^{r-1})
\qquad
\mbox{as}
\qquad
r\rightarrow \infty
\, .
\label{a_b_splitting}
\end{equation}

In the other limiting  case, when $q>>1$ and the spectrum is 
approximately oscillator like, and one finds
\begin{equation}
a \approx 
-2q 
+ 2p\sqrt{q} 
- \frac{(p^2+1)}{2^3}
- \frac{(p^3 + 3p)}{2^7 \sqrt{q}} 
- \frac{(5p^4 + 34p^2+9)}{2^{12} q}
- \frac{(33p^5 + 410p^3 + 405p)}{2^{17} q^{3/2}}
- \cdots
\label{mathieu_limit_low}
\end{equation}
where $p \equiv 2n+1$ which has the clear $(n+1/2)\hbar \omega$ 
oscillator dependence in low order. We will explicitly show,
in Sec.~\ref{sec:low_energy} and 
Appendix~\ref{sec:third_and_4th_appendix},
how this expansion arises using perturbation theory, at least up
to 4th-order, including all of the terms shown above. We note that
handbook expansions exist to sixth-order (in both the low- and high-energy
limits), so that one could improve on the explicit quantum mechanical 
perturbative calculations discussed below by using these results.

While we will focus on approximate perturbative results in the next
two sections, when we do compare to purely numerical results, we will
make use of the following nominal set of parameter values,
\begin{equation}
\hbar = 2\mu = l = 1
\qquad
\quad
\mbox{and}
\qquad
\quad
V_0 = 80
\label{parameters}
\end{equation}
corresponding to $q=160$.  This choice can be seen,  from Fig.~1, to
exhibit a large number of oscillator like states below 
the $+V_0$ threshold. 
For these parameter values, \mathmath \, returns characteristic values with 
the equivalent of double-precision accuracy.

\section{\label{sec:high_energy} High energy rotor limit}

In the limit of vanishing gravitational or electric field, the quantum
pendulum reduces to a free rotor with Schr\"odinger equation 
and solutions given by
\begin{equation}
- \frac{\hbar^2}{2I} \frac{d^2 \psi_{m}(\theta)}{d \theta^2}
= E_{m}^{(0)} \psi_{m}(\theta)
\qquad
\mbox{with}
\qquad
\psi_{m}(\theta) = \frac{1}{\sqrt{2\pi}}e^{im\theta}
\qquad
\mbox{and}
\qquad
E_{m}^{(0)} = \frac{\hbar^2 m^2}{2I}
\label{rotor_zero}
\end{equation}
with $m = 0, \pm 1, \pm 2,...$. The solutions with $+m,-m$ for $|m| \neq 0$
are degenerate, corresponding to the equivalence of clockwise and 
counter clockwise motions. Linear combinations of these solutions can 
be used to obtain eigenfunctions of definite parity as 
\begin{equation}
      \Theta_{m}(\theta) = \left\{ \begin{array}{ll}
               1/\sqrt{2\pi} & \mbox{for $m=0$} \\
               \cos(m\theta)/\sqrt{\pi}& \mbox{for $m>0$} \\
               \sin(m\theta)/\sqrt{\pi} & \mbox{for $m>0$} \\
                                \end{array}
\right.
\, . 
\end{equation}

The classical period associated with this zeroth-order result requires
\begin{equation}
\frac{dE_{m}}{dm} \equiv E_{m}' = \frac{\hbar^2 m}{I}
= \hbar \sqrt{\frac{2E_{m}}{I}}
\end{equation}
so that
\begin{equation}
\tau_{m}^{(0)} \equiv \frac{2\pi \hbar}{|E_{m}'|}
= 2\pi/\sqrt{2E_{m}^{(0)}/I}
\end{equation}
which is consistent with the purely classical result of
Eqn.~(\ref{classical_limit}). The corresponding revival time is 
\begin{equation}
T_{rev} = \frac{2\pi \hbar}{|E_m''/2|} = \frac{4\pi I}{\hbar}
\, . 
\label{rotor_revival}
\end{equation}
These simple results can therefore be used to obtain zeroth- order 
approximations for the
quantized energies, classical periods, and revival times in the 
large-energy rotor limit. We note that in leading order, the superrevival 
and all higher-order correlation times diverge due to the simple quadratic
dependence on quantum number, just as for the one-dimensional infinite well.

Using the standard set of values in Eqn.~(\ref{parameters}),
we first exhibit the zeroth order energy eigenvalue predictions (compared
to the 'numerically exact' results) in Fig.~2, and then compare the
$\tau_{m}^{(0)}$ versus $E_{m}^{(0)}$ values to the classical periodicity 
in Fig.~3 in this same limit. Finally, in Fig.~4, we illustrate the 
predictions for the revival time, $T_{rev}$ versus $E_{m}$. 

The effect of a constant external field (Stark effect) on the spectrum 
of the rotor is a well-known example of second-order degenerate 
perturbation theory \cite{pendulum_martin} - \cite{problem_cronin} 
and the general result for the second-order energy shift is
\begin{equation}
E_{(m)}^{(2)} = \frac{IV_0^2}{\hbar^2 (4m^2-1)}
\label{rotor_second}
\end{equation}
for both even and odd states. Note that the first-order,
and all odd-order shifts vanish for symmetry reasons. The special case 
of $|m|=1$ must be treated separately  with the result that 
the $+1,-1$ states are split to this order 
\cite{pendulum_martin} - \cite{pendulum_sposito},
\cite{problem_kogan}, \cite{problem_cronin}. More generally, it can
be argued \cite{pendulum_khare} that the degeneracy in $a_{r},b_{r}$ is 
split in $r$-th order in perturbation theory, 
as in Eqn.~(\ref{a_b_splitting}).
When this result (Eqn.~(\ref{rotor_second})) is written in terms of
the parameters $a$ and $q$, it reproduces the next-to-leading order of
the 'handbook' expansion in Eqn.~(\ref{mathieu_limit_high})
to ${\cal O}(q^2)$.

This next-to-leading term then provides important corrections not only 
to the energy eigenvalues (especially for lower energy, smaller 
$m$ states), but also to the quantum expressions for 
the classical period (via $|E_{m}'|$) and 
the revival time (via $|E_{m}''|$). Including these next-to-leading order
results gives the second order predictions (shown as boxes) for these
quantities in Figs.~2, 3, and 4. We note that the quantum revival times
decrease from their high-energy value in Eqn.~(\ref{rotor_revival})
as one approaches the critical $E\sim +V_0$ value from above; this is
in contrast to the classical periodicity where the quantum 
$\tau_{m}$ values 
do seem to more closely follow the classical prediction of 
Eqn.~(\ref{high_energy_period}) and exhibit signs of the expected
divergence as $E \rightarrow +V_0$. This term also provides the first
estimate of the superrevival time as one no longer has $E_{m}''' = 0$;
this correction is somewhat akin to the introduction of a finite
superrevival time in the finite well system 
\cite{finite_well}, \cite{other_finite_well} when the well depth is 
not taken to infinity, yielding a non-quadratic dependence on quantum
number.

While we are not aware of any corresponding explicit 4th-order calculation 
of the energy eigenvalues for the 'rotor in a field' system,  one can 
make use of the perturbative expansion obtained by
other methods in Eqn.~(\ref{mathieu_limit_high}) to write the equivalent
of the 4th-order perturbation theory result in the form
\begin{equation}
E_{m}^{(4)} = \frac{V_0^2I^2}{\hbar^6} 
\left[\frac{ 20m^2+7}{(4m^2-1)^3 (4m^2-4)}\right]
\label{rotor_4th}
\end{equation}
and including these additional corrections does yield improved agreement
with both the numerically evaluated energies (in Fig.~2) and in the 
prediction for the energy dependence of the classical period (in Fig.~3): 
it also sharpens the apparent 'dropoff' of the revival time as the energy
approaches the separatrix value (as $E \rightarrow +V_0$ from above) as 
illustrated in Fig.~4. Using the comparisons in Fig.~2 and 3, one can then 
better judge to what extent the
predicted values for the revival times will likely reproduce the actual
values for the complete quantum pendulum problem, as one approaches 
$V_0$ from above. Henceforth, we will use the 4th-order predictions for
the revival (and superrevival) times in comparisons to numerically obtained
values for the pendulum system. We note that the leading (in $m$) terms
of the perturbation results in Eqns.~(\ref{rotor_zero}),
(\ref{rotor_second}), and (\ref{rotor_4th}), as well as higher order
ones in the detailed 'handbook'
expansion in Eqn.~(\ref{mathieu_limit_high}) can be obtained in a very
straightforward manner using the WKB approximation and we outline this
procedure in Appendix~\ref{sec:wkb_appendix}.

As a check on the standard assumption on the relative magnitudes of the
various time scales in such problems, we can evaluate the lowest-order
predictions for $\tau$, $T_{rev}$, and $T_{super}$, using the relations
in Eqn.~(\ref{period_definitions}). The zeroth-order energy in
Eqn.~(\ref{rotor_zero}) gives the leading contributions to the first
two quantities, while the second-order term in Eqn.~(\ref{rotor_second})
gives the leading term for $T_{super}$ and we find the hierarchy
\begin{equation}
T_{super}:T_{rev}:\tau =
(4m^3/q)^2:2m:1
\end{equation}
in the large $m$ limit. We also recall that the next-to-leading corrections
for each of these quantities imply that $\tau$ increases as $E \rightarrow
+V_0$, while both $T_{rev},T_{super}$ decrease so that the time scales
are compressed near the separatrix.

\section{\label{sec:low_energy} Low energy harmonic oscillator limit}

In order to study the low-energy or oscillator-like limit 
of Eqn.~(\ref{basic_equation}), we write
$I = \mu l^2$ and expand the potential energy function for $\theta << 1$,
giving
\begin{equation}
- \frac{\hbar^2}{2\mu l^2} \frac{d^2 \psi(\theta)}{d\theta^2}
- V_0 \left(1 - \frac{\theta^2}{2!} + \frac{\theta^4}{4!}
- \frac{\theta^6}{6!} + \cdots \right) \psi(\theta) = E \psi(\theta)
\, .
\end{equation}
Collecting only the constant and quadratic terms from the potential energy, 
and writing $x \equiv \theta l$ (again, in the spirit of a small angle 
approximation), we can write the exact Hamiltonian in the form
\begin{equation}
\hat{H}_{0} + \hat{H}' 
= \hat{H}_{0} + \sum_{r=2}^{\infty} \hat{H}_{2r}
\equiv
\left[ 
-\frac{\hbar^2}{2\mu} \frac{d^2}{dx^2} + \frac{1}{2} 
\left(\frac{V_0}{l^2}\right) x^2 -V_0
\right]
+ \sum_{r=2}^{\infty} (-1)^{r+1}\frac{V_0 x^{2r}}{(2r)! l^{2r}}
\end{equation}
and we will treat the various $\hat{H}_{2r}$ terms in $\hat{H}'$
as perturbations. We next equate
\begin{equation}
\frac{1}{2} \mu \omega^2 
\Longleftrightarrow 
 \frac{V_0}{2l^2}
\qquad \quad
\mbox{or}
\qquad \quad 
\omega \equiv \sqrt{\frac{V_0}{ml^2}} = \sqrt{V_0/I}
\label{frequency}
\end{equation}
so that the zeroth order energy eigenvalues are trivially given by
\begin{equation}
E_{n}^{(0)} = (n+1/2) \hbar \omega - V_0
= (n+1/2) \hbar \sqrt{V_0/I} - V_0 \, .
\label{sho_zero}
\end{equation}
Expressed in terms of $a$ and $q$, this reduces to 
\begin{equation}
a = -2q + 2(2n+1)\sqrt{q}
\end{equation}
which are the two leading terms of the 'handbook' expansion in 
Eqn.~(\ref{mathieu_limit_low}). This harmonic oscillator approximation
is compared to the numerically obtained quantum pendulum eigenvalues 
in Fig.~2 (as crosses)
as the leading (zeroth) order prediction, for $E<+V_0$.
The classical periodicity from Eqn.~(\ref{period_definitions}) 
is given by
\begin{equation}
\tau_{n} = \frac{2\pi \hbar}{|E'_n|} = \frac{2\pi}{\omega}
\end{equation}
and we superimpose these zeroth order predictions for $(E_n,\tau_n)$
on the classical prediction of Eqn.~(\ref{low_energy_period}) 
in Fig.~5 (as crosses).

The leading correction in this approach comes from treating $\hat{H}_{4}$
using first-order perturbation theory to write
\begin{equation}
E_{n}^{(1)} = \langle n | \hat{H}_{4} | n \rangle
= -\frac{V_0}{4! l^4} \langle n |x^4 | n \rangle
= -\frac{1}{4!} \frac{V_0}{l^4} 
\left[\frac{3\hbar^2}{4\mu^2 \omega^2} (2n^2 + 2n +1) \right]
\end{equation}
using elementary textbook results for the expectation value of $x^4$ in
oscillator states.  As always, the evaluation of the expectation value
of powers of $x$ is most easily done using raising and lower operator
formalism, with
\begin{equation}
x = \sqrt{\frac{\hbar}{2\mu \omega}}
\left( \hat{A} + \hat{A}^{\dagger}\right)
\end{equation}
where
\begin{equation}
\hat{A} |n \rangle = \sqrt{n} | n -1 \rangle
\qquad
\quad
\mbox{and}
\qquad
\quad
\hat{A}^{\dagger}  |n \rangle = \sqrt{n+1} |n +1 \rangle
\end{equation}
and we use this method in all our calculations below. 
Using the identification in Eqn.~(\ref{frequency}),
we find that this reduces to
\begin{equation}
E_{n}^{(1)} = - \frac{\hbar^2}{32I}(2n^2 + 2n +1)
\, .
\label{sho_first}
\end{equation}
In terms of $a,q$, this contributes a $V_0$-independent 
correction of the form 
\begin{equation}
a^{(1)} = - \frac{(4n^2 +4n +2)}{8} = -\frac{(p^2+1)}{8}
\end{equation}
where $p=2n+1$, which is the $q$-independent term in 
Eqn.~(\ref{mathieu_limit_low}).

This correction, arising from the anharmonicity of the potential, 
not only provides an improved prediction for the
energy-period relationship in Fig.~5, but is also important as it 
gives the lowest-order prediction for the revival period in the 
low-energy limit,  as in Refs.~\cite{morse} and \cite{carpets}. 
The revival time is given by 
\begin{equation}
T_{rev} = \frac{2\pi \hbar}{|E_n''/2|}
= \frac{32\pi I}{\hbar}
\label{sho_revival}
\end{equation}
which is independent of $V_0$ and $8$ times the lowest-order result for
the revival time for the pure rotor system from Eqn.~(\ref{rotor_revival}).
Thus, the revival time in the low-energy limit is predicted in terms of 
physical parameters of the systems, by the very specific anharmonicity 
arising from the form of the $\cos(\theta)$ potential.

We show this first non-vanishing prediction for the revival time in the
low-energy, oscillator like limit in Fig.~6 (as diamonds.) At this order,
the superrevival time is formally infinite since the spectrum is still 
approximately quadratic in $n$.

At next order, one obtains two contributions, one from including
the effect of $\hat{H}_{6}$ in first order perturbation theory, and
a second from using $\hat{H}_{4}$ in second-order. The results can
be written in the forms
\begin{equation}
E_{n}^{(2,a)} \equiv \langle n |\hat{H}_{6} | n \rangle
= +\frac{V_0}{6! l^6} \langle n| x^6 | n \rangle 
= \frac{V_0}{6! l^6} \left(\frac{\hbar}{2\mu \omega}\right)^3
5(4n^3 + 6n^2 + 8n+3)
\end{equation}
and
\begin{eqnarray}
E_{n}^{(2,b)} & = & \sum_{j\neq n} 
\frac{\langle n| \hat{H}_{4} |j\rangle
\langle j | \hat{H}_{4}| n \rangle}{(E_{n}^{(0)} - E_{j}^{(0)})}
 \nonumber \\
&  = & \left(-\frac{V_0}{4!l^4}\right)^2 \frac{1}{\hbar \omega}
\sum_{j\neq n} \frac{\langle n|x^4|j \rangle 
\langle j|x^4|n \rangle}{(n-j)}
\\
& = & 
\left(-\frac{V_0}{4!l^4}\right)^2 \left[\frac{-2}{\hbar \omega}\right]
\left(\frac{\hbar}{2\mu \omega}\right)^4
(34n^3 + 51n^2 + 59n +21) \nonumber 
\end{eqnarray}

Again, using $\omega = \sqrt{V_0/I}$ we then find the total second-order
contribution to be
\begin{equation}
E_{n}^{(2)} =  E_{n}^{(2,a)} + E_{n}^{(2,b)} 
= -\frac{\hbar^3}{I^2} \sqrt{\frac{I}{V_0}} 
\frac{(2n^3+3n^2 + 3n+1)}{512}
\label{sho_second}
\end{equation}	
which agrees (when expressed in terms of $a$ and $q$) with the expansion 
in Eqn.~(\ref{mathieu_limit_low}) and
improves agreement with the 'exact' (numerically evaluated) pendulum values 
as shown in Fig.~2.
The $(E_n,\tau_n)$ values including these corrections are also
plotted in Fig.~5 and similarly show improved agreement in the $E<+V_0$
regime. In analogy to the high-energy rotor case, the lowest-order
prediction of a fixed value of the revival time in this limit,
in Eqn.~(\ref{sho_revival}) is changed in the next order so that
$T_{rev}$ decreases as $E \rightarrow +V_0$ from below.

We can proceed in an identical (if increasingly more difficult) fashion 
to find the 3rd and 4th order corrections, namely
\begin{eqnarray}
E_{n}^{(3)} & = & - \frac{\hbar^4}{V_0 I^2} 
\frac{(5n^4+10n^3+16n^2+11n+3)}{8192} 
\label{sho_third} \\ 
E_{n}^{(4)} & = & - \frac{\hbar^5}{I^{5/2}V_0^{3/2}}
\frac{(66n^5 + 165n^4+370n^3+390n^2+225n+53)}{2^{19}}
\label{sho_4th} 
\end{eqnarray}
and some of the details of their calculation are presented in 
Appendix~\ref{sec:third_and_4th_appendix}. In each case, 
the results match the 'handbook' expansion
from Eqn.~(\ref{mathieu_limit_low}); they also improve agreement with 
numerically obtained values (as in Fig.~2) and with the classical 
periodicity (in Fig.~5), as well as steepening the 'dropoff' of the 
predicted quantum revival time in this limit as shown in Fig.~6. 
We see that the quantum revival times decrease (at least initially) 
as one goes up in energy,  in contrast to the quantum periods 
($\tau_{n}$) which increase in accordance (again, initially) with 
the classical prediction as one approaches the classical divergence 
in $\tau$.

We can also check the hierarchy of time scales, as done at the
end of Sec.~\ref{sec:high_energy}, in the low-energy limit. In
this case, the lowest-order predictions for $\tau$, $T_{rev}$,
and $T_{super}$ come from Eqns.~(\ref{sho_zero}),
(\ref{sho_first}), and (\ref{sho_second}) respectively and give
the simple ratios
\begin{equation}
T_{super}:T_{rev}:\tau =
(8\sqrt{q})^2: (8\sqrt{q}):1
\end{equation}
and higher-order corrections lead to increasing (decreasing)
values of $\tau$ ($T_{rev},T_{super}$) as $E \rightarrow +V_0$
as in the high-energy rotor limit, with the various time scales
being 'compressed' near the separatrix.

\section{\label{sec:results} Classical periodicity, revival and 
superrevival times for the pendulum}

We have derived perturbative expressions for the quantized
energies in both the low- and high-energy limits, and can therefore
evaluate  approximate expressions for all of the relevant time scales
in Eqn.~(\ref{period_definitions}), as well as judging their
relevant range of validity by comparing to 'exact' values as in
Fig.~1. In order to examine the behavior of the classical periods,
and revival and superrevival times,  near the separatrix (for $E \approx
+V_0$) where neither approximation scheme is applicable, 
we need to use numerically obtained values. With the energy
spectrum, $E_{n}$ versus $n$, obtained in this way, we can approximate
the required derivatives with respect to quantum number by making
associations such as 
\begin{equation}
\frac{dE_{n}}{dn} \Longrightarrow 
\frac{\Delta E_{n}}{\Delta n} = E_{n+1} - E_{n}
\qquad
\mbox{and}
\qquad
\frac{d^2 E_{n}}{dn^2} \Longrightarrow
\frac{\Delta^{(2)} E_{n}}{\Delta n^2}
= E_{n+1} - 2E_{n} + E_{n-1}
\label{discrete}
\end{equation}
with similar expressions for higher order derivatives. However, because
of the degeneracy between even and odd rotor-like states in the 
high energy
limit, if we simply enumerate all of the allowed states in some standard
ordering, we find vanishing 'derivatives' in the free rotor limit which
are artifacts. In the construction of a general, high-energy wave packet,
one would find approximately equal contributions from both the
(almost exactly degenerate) even ($ce_{2m}$) and odd ($se_{2m}$) 
solutions
which would be only 'counted' once. 
To avoid this difficulty, we can classify the states by their parity 
and evaluate the
approximate derivatives in Eqn.~(\ref{discrete}) separately in both the
even and odd cases.
For the low-energy, oscillator limit, this means that we 'sample' every
other state and so, for example, obtain first-derivative type terms
which are a factor of two too small (since effectively we have used
$\Delta n = 2$ instead of $\Delta n = 1$), with a similar factor of $4$
for the second derivative term. This procedure is not as artificial
as it may seem, since one can imagine constructing wave packets of
definite parity, even or odd, at appropriate points of symmetry for the
pendulum (either at $\theta =0$ or $\theta = \pi$) which would then
have contributions only from the even or odd parity eigensolutions.

Following this procedure, we construct approximations for the 
first derivatives by generalizing Eqn.~(\ref{discrete}) slightly by
associating the 'average' energy
\begin{equation}
\overline{E_{n}} = \frac{1}{2} (E_{n+1} + E_{n})
\qquad
\mbox{with}
\qquad
\frac{\Delta E_{n}}{\Delta n} = E_{n+1} - E_{n} \equiv \Delta E_{n}
\end{equation}
so that we associate 
\begin{equation}
\overline{E_{n}}
\qquad
\mbox{and}
\qquad
\overline{\tau}_{n} = \frac{2\pi \hbar}{|\Delta E_{n}|}
\, . 
\end{equation}
We plot, in Fig.~7, both the even-parity (crosses) 
and odd-parity (diamonds) 
values for $(\overline{E}_n,\overline{\tau}_{n}$) obtained in this way.
(Note that the values for $E<+V_0$ are shown on an appropriately scaled
(factor of $2$) axis, compared to those values for $E> +V_0$.)
We note that the 'exact' data do closely follow the classical curves from
Eqns.~(\ref{low_energy_period}) and (\ref{high_energy_period}), and do
exhibit a large, but finite peak at $E \approx +V_0$, in both sets of
data. The lack of a true divergence is understandable given the
quantized nature of the energy eigenvalues and has been discussed
in terms of the explicit time evolution of wave packet solutions
as described in Ref.~\cite{pendulum_cook}, which also outlines
simple uncertainty principle arguments.

Turning now to the question of the behavior of the revival time
as $E \rightarrow +V_0$ (from either above or below), while there is
no classical prediction comparable to that for $\tau(E)$ versus $E$,
we can make use of the 4th-order perturbative results to evaluate
$T_{rev} = 2\pi \hbar/|E_{n}''/2|$ versus $E_{n}$ which will give
good representations in both the low- and high-energy limits. Using
the 4th-order results plotted in Figs.~3 and 5, we produce the
limiting predictions for $T_{rev}$ versus $E_{n}$, shown as solid curves, 
in Fig.~8. The numerically obtained 'exact' results, using the
prescription in Eqn.~(\ref{discrete}), are also shown in Fig.~8,
again with the low-energy 'data' ($E< +V_0$) shown scaled,
now by a factor of $4$. Once again, the agreement with the 
perturbative predictions
in the two limits is good, with $T_{rev}$ clearly decreasing from the
appropriate $E<<+V_0$ and $E>>+V_0$ limits until very near $E \approx
+V_0$ where it reaches a large, but finite value. Similar results can 
be seen for the superrevival time, $T_{super}$, comparing the predictions
from 4th-order perturbation theory in both the low- and high-energy
limits with discretized derivatives of the numerically 
obtained results (appropriately scaled for $E<+V_0$),
yielding plots which are quite similar to Fig.~8.

\section{\label{sec:conclusions} Discussion and conclusions}

We have examined the quantized energy eigenvalue spectrum of the
quantum pendulum problem, focusing on information about the
classical periodicity and quantum revivals and superrevivals. Using
4th-order perturbative approximations for the low-energy (oscillator) 
 and high-energy (rotor) limits of the pendulum, we have found good
agreement with numerically obtained eigenvalues. This has allowed
us to derive expressions for the revival and superrevival times in
these limits, as well as to establish limits on their validity.
The revival times for the pendulum decrease from constant values,
proportional to $\hbar/I$, as one approaches the separatrix from
both above and below. Near the separatrix, we have used numerically
obtained values to show that the classical period exhibits a large,
but finite value at $E \approx +V_0$, while the revival times
also increase (very near this value).

This problem is an interesting example of a quantum system which exhibits
a rich variety of classical periodicities and quantum revival and
superrevival times, all of which are perturbatively calculable in limiting 
cases, as well as classically divergent behavior at a special value, which 
is 'softened' by quantum effects. The behavior of the time scales as
they near the critical value at the separatrix, where they can be
clearly seen to compress towards each other as $E \rightarrow +V_0$,
differs from the pattern seen in other systems with quantum revivals.
One can speculate on possible experimental realizations of such behavior
in systems exhibiting hindered internal rotations 
\cite{hindered_book1} - \cite{pendulum_baker}.

The energy level structure of the quantum pendulum is, to a large extent,
determined by the numerical (and dimensionless) value of $q$ 
and we can discuss 
order-of-magnitudes estimates for $q$ for some model systems, and how they 
scale with the physical parameters involved.
In the classical limit of a macroscopic pendulum system under the influence 
of gravity, with $m \sim 1\,kg$ and $l \sim 1\,m$, we find $q \sim 10^{70}$
so that there is the expected huge number of oscillatory states below 
the separatrix and one is clearly in the correspondence principle limit.
 At the other extreme, for an electron $(m_e$) restricted 
to a radius characteristic of that of a carbon nanotube ($L \sim 1\,nm$), 
under the influence of gravity, we find that $q \sim 10^{-18}$ and the 
system is automatically in the high-energy or free-rotor limit. In this
case, the revival time should be given to an excellent approximation by
Eqn.~(\ref{rotor_revival}), with no corrections due to the external potential,
as well as having $T_{super} \rightarrow \infty$. For the 
more interesting case of an electron (or ion) subject to an electric field 
of order $E_{0} = 100\,V/m$,  we find 
$q \sim 10^{-6}(M/m_e)(E/E_0)(l/1\, nm)^3$ so that for electrons in
such fields the high-energy free rotor limit is also applicable. 
On the other hand, for ions ($M/m_e > 10^4$) in strong fields 
($E > 10^5\,V/m$) in larger diameter ($L > 10\,nm$) geometries,  one
has $q \sim 10^4$ and an interesting number of vibrational states could be
present. This range is
perhaps more typical of that explored in studies of hindered internal
rotations \cite{hindered_book1} -\cite{hindered_ercolani}.
 We have found that the revival times, in both the high-energy
(Eqn.~(\ref{rotor_revival})) and low-energy 
(Eqn.~(\ref{sho_revival})) limits scale as $T_{rev} \propto I/\hbar$, which
gives numerical values in the range,
$T_{rev} \sim (100\,fs - 1000\, fs) \, (M/m_e)(l/1\,nm)^2$, 
independent of the pendulum potential, $V_0$.

A simple extension of this system would be to consider a particle confined
to a circular (radius $l$) cylinder  of length $L$, subject to a field 
perpendicular to the central axis. This separable system would then have 
quantized energies consisting of the quantum pendulum values discussed
here plus a term given by $E^{(z)}_{k} = \hbar^2 k^2/2\mu L^2$ corresponding
to the quantized motion along the cylinder. The pattern of classical 
periodicities and quantum revivals in systems characterized by two quantum 
numbers has been discussed \cite{bluhm_2d}, \cite{other_2d}. A more
careful study of this case might be motivated by its possible realization
in the geometry of carbon nanotubes. A related two-dimensional 
generalization would be to consider the point object restricted 
to the surface of a sphere which has a richer set of possible classical
motions.

\begin{acknowledgments}
This work was supported in part by the National Science Foundation
under Grant DUE-9950702.  One of us (R. W. R.) also thanks H. Kaplan 
for conversations regarding aspects of this problem as well as the 
organizers of the 2002 Gordon Conference on Physics Research and 
Education (Quantum Mechanics) for their hospitality.
\end{acknowledgments}

\appendix

\section{\label{sec:wkb_appendix} WKB approximation solutions 
to the Mathieu equation}

We briefly describe in this Appendix a simple WKB-based approximation
which gives the leading order (in $m$) results for the (approximately
degenerate) characteristic values, $a_{r},b_{r}$, in the high-energy 
limit (rotor-like) of the quantum pendulum problem, as an expansion in 
powers of $V_0$. In this context, it reproduces the leading $m$ behavior
of well-known 'handbook' results in Refs.~\cite{mac_mathieu} and
\cite{stegun} for the characteristic values $a_{m},b_{m}$ in terms of the
parameter $q$, in a very straightforward way. 

We first rewrite the classical energy conservation connection
\begin{equation}
\frac{p_{\theta}^2}{2I} - V_0\cos(\theta) = E
\end{equation}
in the form
\begin{equation}
p_{\theta} = \sqrt{2I} \sqrt{E + V_0\cos(\theta)}
\, . 
\end{equation}
In the spirit of a WKB approach, we quantize the integral of 
$p_{\theta}$
in the form
\begin{equation}
\sqrt{2I}\int_{-\pi}^{+\pi} \sqrt{E + V_0\cos(\theta)}\,d\theta
= 
\int_{-\pi}^{+\pi}\, p_{\theta}\,d\theta = m\pi \hbar
\, . 
\label{wkb_approx}
\end{equation}
For the physical situation of the quantum pendulum, we would restrict
ourselves to even values of $m$, but for comparison to the general result
inspired by Mathieu equations, we allow any integral $m$ values. In terms
of the Mathieu equation parameters in Eqn.~(\ref{mathieu_parameters})
we can write this in the form
\begin{equation}
2m\pi = \sqrt{a} \int_{-\pi}^{+\pi} 
\left(1 + \frac{2q}{a}\cos(\theta)\right)^{1/2} \,d\theta
\end{equation}
One can expand the integrand as a series in $q/a$ and perform the
angular integrals, leaving only the even terms present, to obtain
\begin{equation}
m  = \sqrt{a}\left(1 - \frac{q^2}{4a^2} - \frac{15q^4}{64a^2}
- \frac{105q^6}{256a^6} + \cdots \right)
\end{equation}
or
\begin{equation}
a = m^2\left(1 - \frac{q^2}{4a^2} - \frac{15q^4}{64a^2}
- \frac{105q^6}{256a^6} + \cdots \right)^{-2} \, . 
\end{equation}
This can be solved recursively in powers of $q^2$ to obtain the
approximation solution
\begin{equation}
a_{m} = m^2 + \frac{q^2}{2m^2} + \frac{5q^4}{32m^6} +
\frac{9q^6}{64m^{10}} + \cdots 
\,\, .
\end{equation}
and this form agrees with the handbook results of 
Refs.~\cite{mac_mathieu} and \cite{stegun}. One can then easily extend 
this result (again to leading order in $m$) to quite high order.

\section{\label{sec:third_and_4th_appendix} Third- and fourth-order 
perturbation theory results in the oscillator limit}

For the third-order correction to the low-energy, harmonic oscillator
limit, we need to evaluate three distinct contributions, namely,
(i) the  $\hat{H}_{8}$ term in first order,
(ii) the $\hat{H}_{6}$ and $\hat{H}_{4}$ terms (in both orders)
in second order, and
(iii) the $\hat{H}_{4}$, $\hat{H}_{4}$, $\hat{H}_{4}$ terms in
third order.  For example, the first contribution is
\begin{equation}
- \frac{V_0}{l^8 8!} \langle n | x^8 | n \rangle
= - \left[\frac{\hbar^2}{I^2V_0}\right]
\frac{5(14n^4+28n^3+70n^2+56n+21)}{8!2^4}
\end{equation}
The expression for the third-order energy shift is
presented in some elementary texts, but we reproduce it below for
completeness, specifically
\begin{equation}
E_{n}^{(3)}  =  \sum_{j\neq n} \sum_{l\neq n}
\frac{
\langle n|\hat{H}'|j \rangle
\langle j|\hat{H}'|l \rangle
\langle l|\hat{H}'|n \rangle
}{
(E_{n}^{(0)} - E_{j}^{(0)})
(E_{n}^{(0)} - E_{l}^{(0)})
}
-
\langle n|\hat{H}'|n \rangle
\sum_{j\neq n}
\frac{
\langle n|\hat{H}'|j \rangle
\langle j|\hat{H}'|n \rangle
}{
(E_{n}^{(0)} - E_{j}^{(0)})^2
}
\label{third_order}
\end{equation}
Including all three sets of terms, we obtain the result shown
in Eqn.~(\ref{sho_third}).

For the increasingly complex matrix algebra involved in the evaluation
of such terms, we make use of symbolic manipulation programs 
(in our case \mathmath) and write
\begin{equation}
\langle i | x | j \rangle
= \sqrt{\frac{\hbar}{2\mu \omega}}
\langle i | \hat{A} + \hat{A}^{\dagger} | j \rangle
= \sqrt{\frac{\hbar}{2\mu \omega}} \left[\sqrt{j}\delta_{i,j-1}
+ \sqrt{j+1}\delta_{i,j+1} \right]
\end{equation}
and evaluate the required matrix elements of higher powers 
by repeated matrix multiplication. We note that in third-order
(and beyond) in perturbation theory, there are consistently terms
including factors of the form  $\langle n| \hat{H}'|n \rangle$. 
At third-order, and beyond, in intermediate steps of the calculation, 
the individual expressions one obtains may be of higher order (in $n$) 
than are present in the final result, and the presence of such 
cancellations can be used as a check on the internal consistency of 
the calculation.

Finally, for the 4th-order terms, we require the contributions from
(i) $\hat{H}_{10}$ in first order,
(ii) $\hat{H}_{8},\hat{H}_{4}$ (both orderings) in second order,
(iii) $\hat{H}_{6},\hat{H}_{6}$,  also in second order,
(iv), $\hat{H}_{6},\hat{H}_{4},\hat{H}_{4}$ (three orderings) in
third order, and
(v) $\hat{H}_{4},\hat{H}_{4},\hat{H}_{4},\hat{H}_{4}$ in 4th-order.
The explicit expression for the energy shift in 4th order, in terms
of unperturbed wavefunctions and energies, is less often seen 
(see, {\it e.g.}, \cite{epstein}) so we also reproduce it here for 
completeness.
\begin{eqnarray}
E_{n}^{(4)} & = & \sum_{j\neq n} \sum_{l\neq n} \sum_{r\neq n}
\frac{
\langle n|\hat{H}'|j \rangle
\langle j|\hat{H}'|l \rangle
\langle l|\hat{H}'|r \rangle
\langle r|\hat{H}'|n \rangle
}{
(E_{n}^{(0)} - E_{j}^{(0)})
(E_{n}^{(0)} - E_{l}^{(0)})
(E_{n}^{(0)} - E_{r}^{(0)})
}
\nonumber \\
& & 
-
\langle n| \hat{H}'| n \rangle
\sum_{j\neq n} \sum_{l\neq n} 
\frac{
\langle n|\hat{H}'| j \rangle
\langle j|\hat{H}'| l \rangle
\langle l|\hat{H}'| n \rangle
}{
(E_{n}^{(0)} - E_{j}^{(0)})
(E_{n}^{(0)} - E_{l}^{(0)})^2
}
\nonumber \\
& & 
-
\langle n| \hat{H}'| n \rangle
\sum_{j\neq n} \sum_{l\neq n} 
\frac{
\langle n|\hat{H}'| j \rangle
\langle j|\hat{H}'| l \rangle
\langle l|\hat{H}'| n \rangle
}{
(E_{n}^{(0)} - E_{j}^{(0)})^2
(E_{n}^{(0)} - E_{l}^{(0)})
}
\label{4th_order}\\ 
& &
+
\langle n|\hat{H}' | n \rangle 
\langle n|\hat{H}' | n \rangle 
\sum_{j\neq n}
\frac{
\langle n|\hat{H}'|j \rangle
\langle j|\hat{H}'|n \rangle
}{
(E_{n}^{(0)} - E_{j}^{(0)})^3
}
\nonumber \\
& &
-
\left[
\sum_{r\neq n}
\frac{
\langle n|\hat{H}'|r \rangle
\langle r|\hat{H}'|n \rangle
}{
(E_{n}^{(0)} - E_{r}^{(0)})
}
\right]
\left[
\sum_{j\neq n}
\frac{
\langle n|\hat{H}'|j \rangle
\langle j|\hat{H}'|n \rangle
}{
(E_{n}^{(0)} - E_{j}^{(0)})^2
}
\right]
\nonumber
\end{eqnarray}

As a cross-check of our more automated approach to the 3rd and 4th
order calculations, we have followed the same approach for the
1st- and 2nd-order results and confirm that we reproduce the values
in Eqns.~(\ref{sho_first}) and (\ref{sho_second}). 
Perhaps more interestingly, 
we also note that there are three simple extensions of the 'basic' 
oscillator, defined by the zeroth-order 
potential $V(x) = \mu \omega^2 x^2/2$, which can be used as 
benchmarks
for such higher order calculations. First of all, the addition 
of a constant perturbing potential term, $V'_{0}$, should only 
affect the energy spectrum through the first-order correction 
$E_{n}^{(1)} = \langle n | V'_{0} | n\rangle = V'_{0}$ 
and one can easily check 
that all higher order corrections vanish in this case.

The inclusion of a linear term of the form $V'(x) = -Fx$ can be solved
exactly by completing the square giving
\begin{equation}
V(x) = \frac{1}{2} \mu \omega^2 x^2 - Fx
= \frac{1}{2} \mu \omega^2 \left(x - \frac{F}{\mu \omega^2}\right)^2
- \frac{F^2}{2\mu \omega^2}
= V(x-x_0) - \tilde{E}_0
\end{equation}
to give $E_{n}' = E_{n} - F^2/2\mu/\omega^2 = E_{n} - \tilde{E}_0$. 
Thus, the inclusion of an $\hat{H}' = -Fx$ perturbation should find a 
non-vanishing result only for $E_{n}^{(2)}$ and we confirm this. 

Finally, the inclusion of a small quadratic correction to the original
oscillator potential by $V'(x) = \lambda x^2$ leads to a simple 
redefinition of the oscillator frequency 
$\omega' = \omega \sqrt{1+2\lambda/\hbar\omega^2}$ and the result
\begin{equation}
E_{n}' = (n+1/2) \hbar \omega' = E_{n} \left(1 + \frac{2 \lambda}{\hbar
\omega^2}\right)^{1/2}
\end{equation}
which can be expanded as an ordinary power series in $\lambda$. One can
then  compare this result to that obtained by evaluation of the
first- through 4th-order perturbation results, including those of
Eqns.~(\ref{third_order}) and (\ref{4th_order}) and we find complete 
agreement.

\newpage

\begin{figure}
\epsfig{file=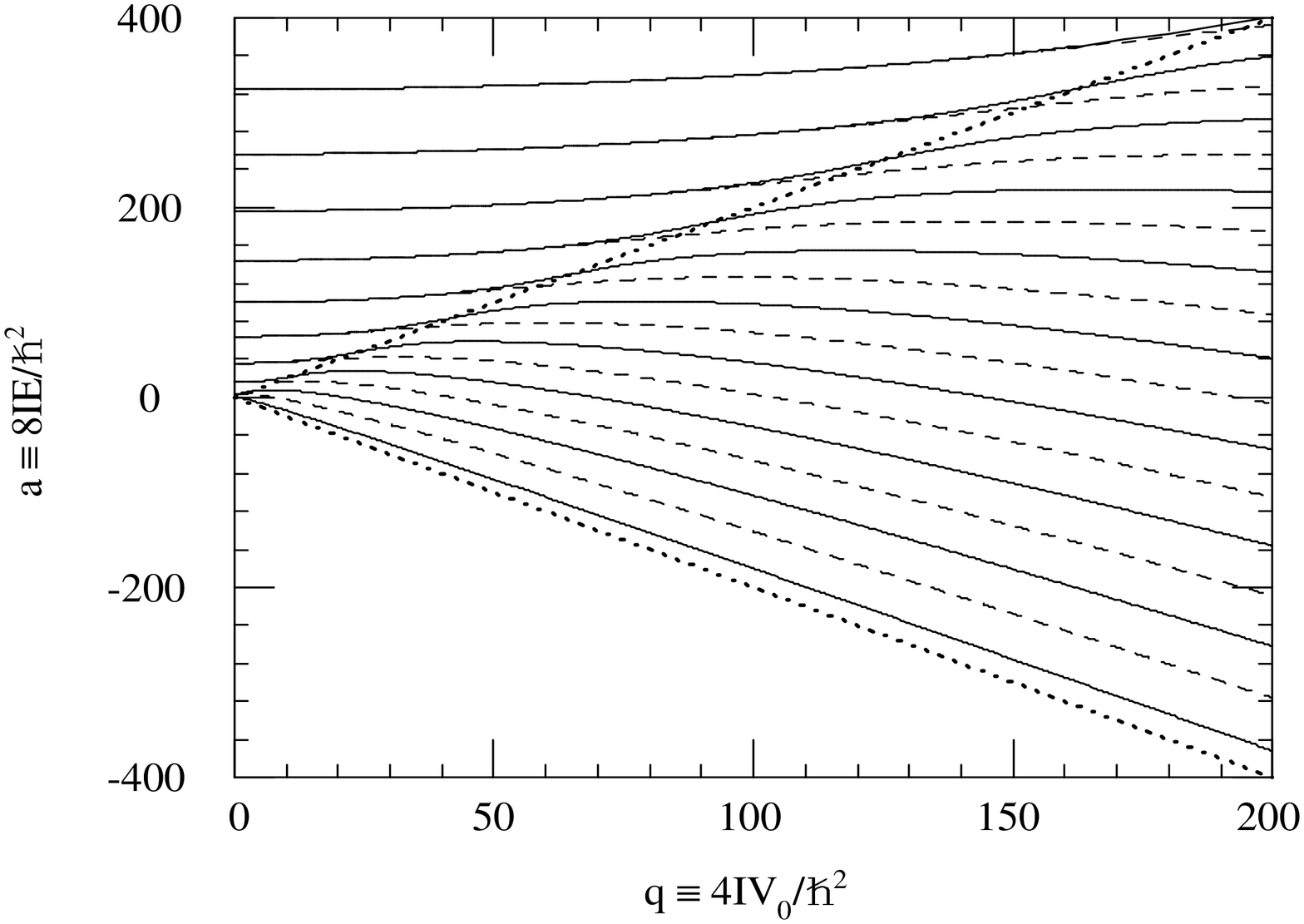,width=9cm,angle=270}
\caption{Plots of the characteristic values for the Mathieu equation,
$a_{2m}$ (for even solutions, solid curves) and $b_{2m}$ 
(for odd solutions,
dashed curves) versus $q$ for the quantum pendulum. The dotted lines
correspond to $a=\pm 2q$ or $E = \pm V_0$. Note that for a given
value of $q$ (at least for $a<<q$) that the gap between the 
lowest energy state (lowest
solid curve) is roughly one-half of the spacing between solid and
dashed curves, corresponding to the zero-point energy in the 
oscillator limit.
}
\end{figure}

\newpage

\begin{figure}
\epsfig{file=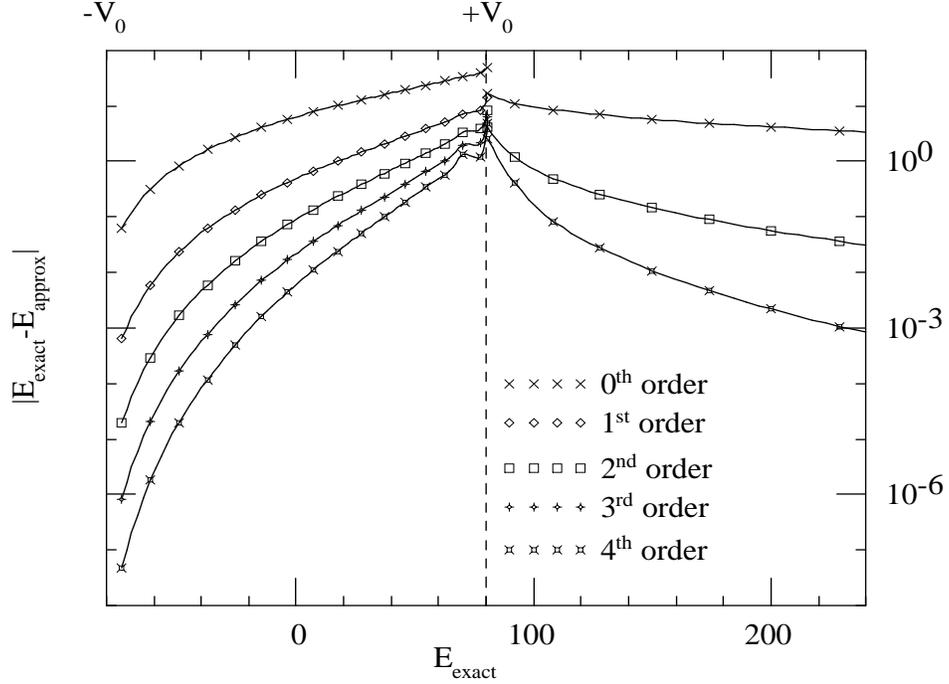,width=9cm,angle=270}
\caption{Plots of the difference between the perturbative approximations
for the energy eigenvalues and the 'exact' values (from direct numerical
evaluation) in various orders of perturbation theory. The parameter
values in Eqn.~(\ref{parameters}) are used. For the $E>+V_0 = 80$
cases (using the high energy rotor limit) the zeroth, second, and 4th-order 
expressions in Eqns.~(\ref{rotor_zero}), (\ref{rotor_second}), 
and (\ref{rotor_4th}) are used. For the $E<+V_0$ case, the low-energy
oscillator approximations in Eqns.~(\ref{sho_zero}), (\ref{sho_first}),
(\ref{sho_second}), (\ref{sho_third}), and (\ref{sho_4th}) are used.}
\end{figure}

\newpage

\begin{figure}
\epsfig{file=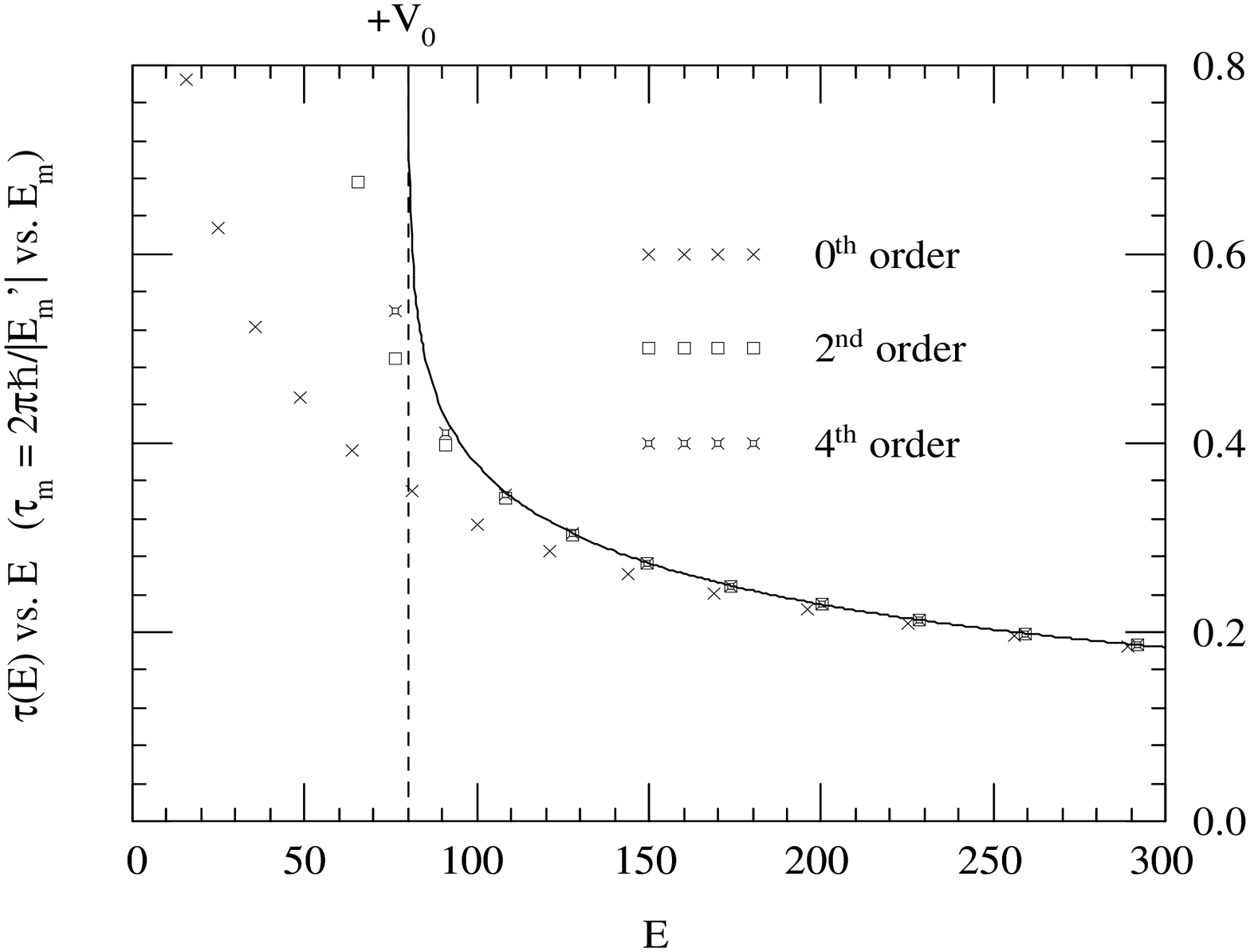,width=9cm,angle=270}
\caption{Plots of $\tau(E)$ versus $E$ for the classical 
pendulum from Eqn.~(\ref{high_energy_period}) (solid curve) and of the 
equivalent quantum values, $\tau_{m} \equiv  2\pi \hbar/|E_m'|$ 
versus $E_{m}$  using 
zeroth-        (from Eqn.~(\ref{rotor_zero}), as crosses), 
second-        (from Eqn.~(\ref{rotor_second}), as squares),
and 4th-order (from Eqn.~(\ref{rotor_4th}), as stars) 
perturbation theory.}
\end{figure}

\newpage

\begin{figure}
\epsfig{file=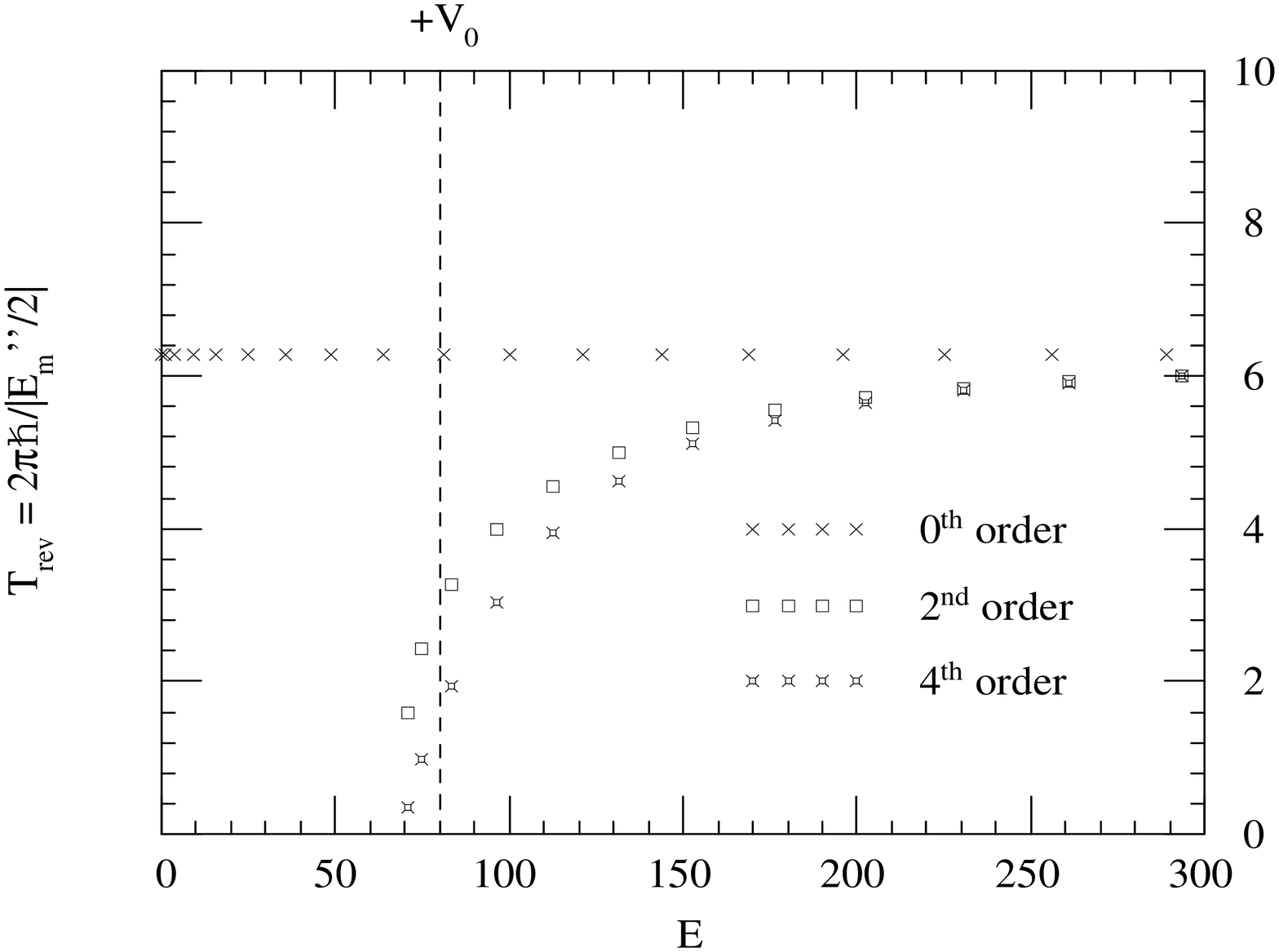,width=9cm,angle=270}
\caption{Plots of the revival time $T_{rev}$ versus $E$ using
$T_{rev} \equiv 2\pi\hbar/|E_{m}''/2|$ in zeroth-, second-, 
and 4th-order
perturbation theory (with the same notation as Fig.~3.)}
\end{figure}

\newpage

\begin{figure}
\epsfig{file=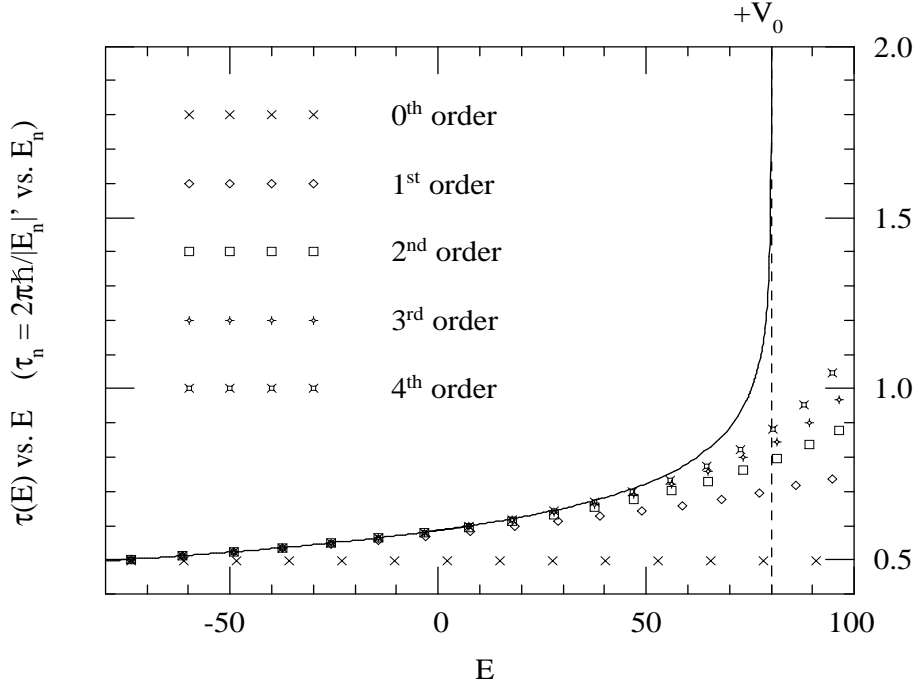,width=9cm,angle=270}
\caption{Plots of $\tau(E)$ versus $E$ for the classical pendulum
from Eqn.~(\ref{low_energy_period}) (solid curve) and of the equivalent
quantum values, $\tau_{m} = 2\pi \hbar/|E_m'|$ versus $E_{m}$ 
using 
zeroth (from Eqn.~(\ref{sho_zero}), as crosses), 
first (from Eqn.~(\ref{sho_first}), as diamonds),
second (from Eqn.~(\ref{sho_second}), as squares),
third (from Eqn.~(\ref{sho_third}), as bursts), 
and 4th order (from Eqn.~(\ref{rotor_4th}), as stars) 
perturbation theory.}
\end{figure}

\newpage

\begin{figure}
\epsfig{file=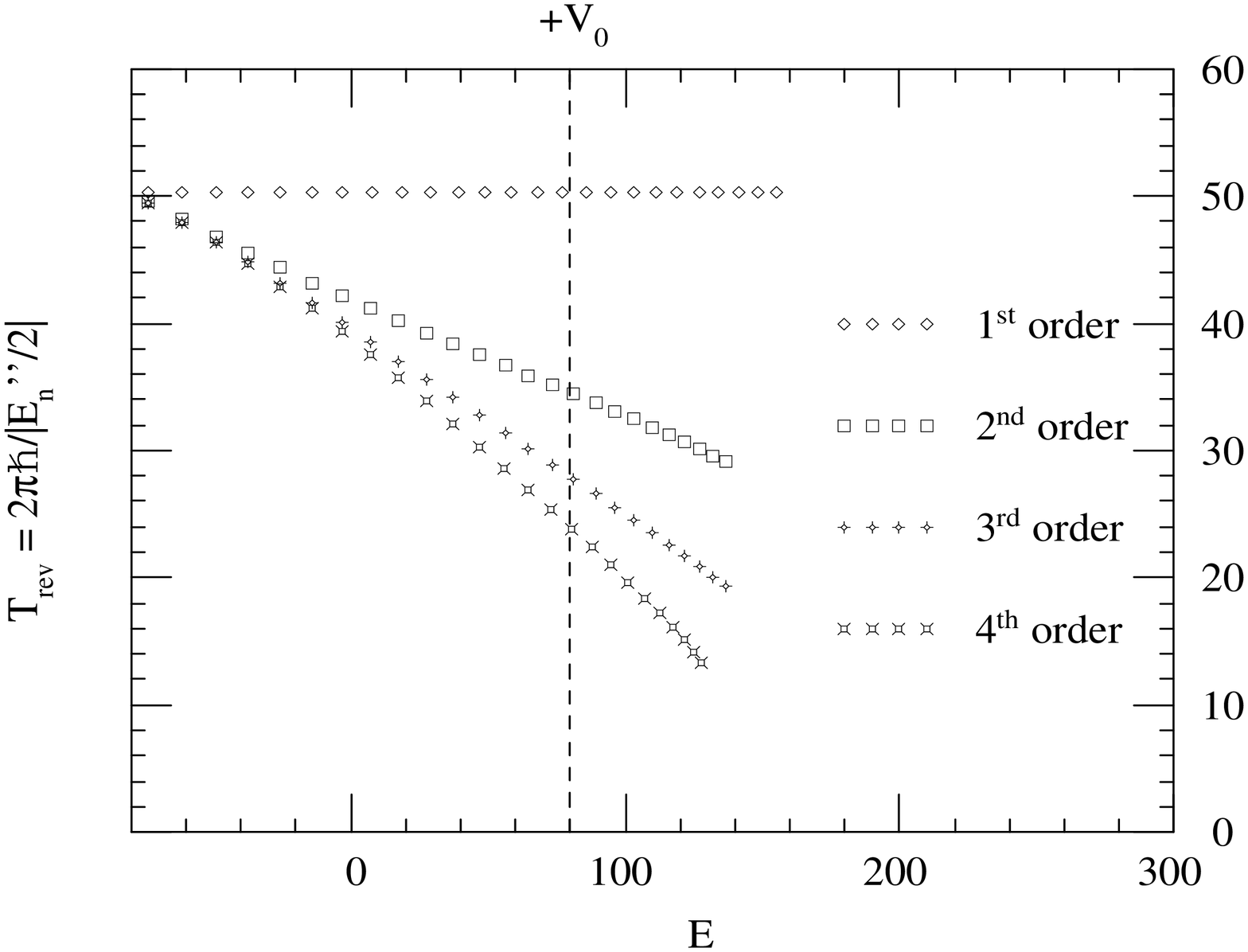,width=9cm,angle=270}
\caption{Plots of the revival time $T_{rev}$ versus $E$ using
$T_{rev} \equiv 2\pi\hbar/|E_{m}''/2|$ in first-, second-, third-, 
and 4th-order perturbation theory (with the same notation as Fig.~5.)}
\end{figure}

\newpage

\begin{figure}
\epsfig{file=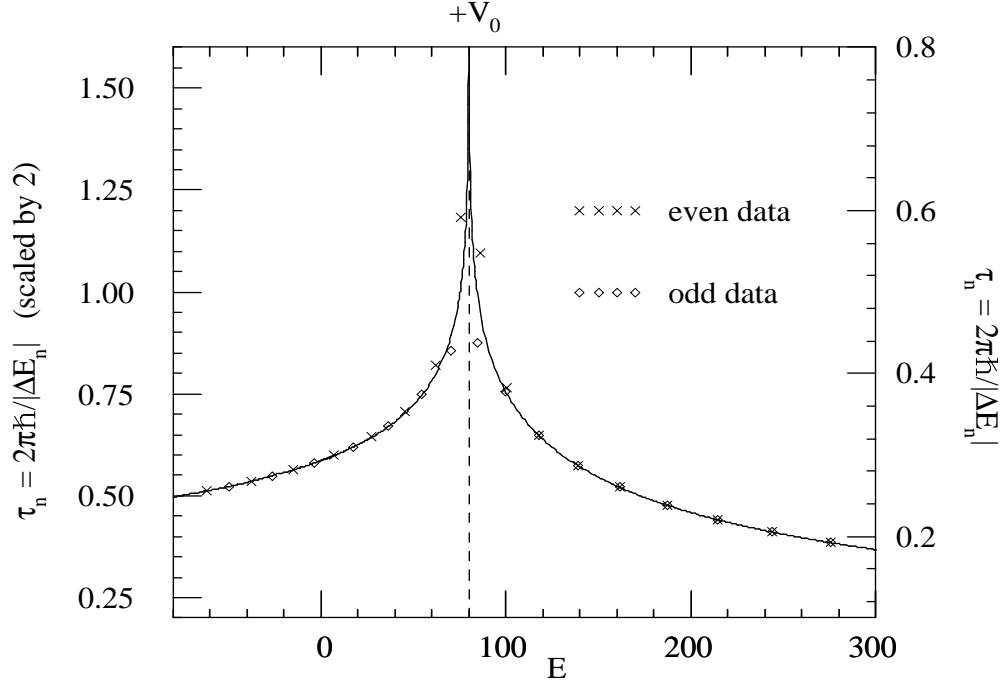,width=9cm,angle=270}
\caption{Plot of $\tau_n \equiv 2\pi \hbar/|\Delta E_n|$ versus
$E_n$ for the parameters in Eqn.~(\ref{parameters}) for even (crosses)
and odd (diamonds) energy eigenvalues. The classical expressions
for $\tau(E)$ versus $E$ from Eqns.~(\ref{low_energy_period} and
(\ref{high_energy_period}) are used. }
\end{figure}

\newpage

\begin{figure}
\epsfig{file=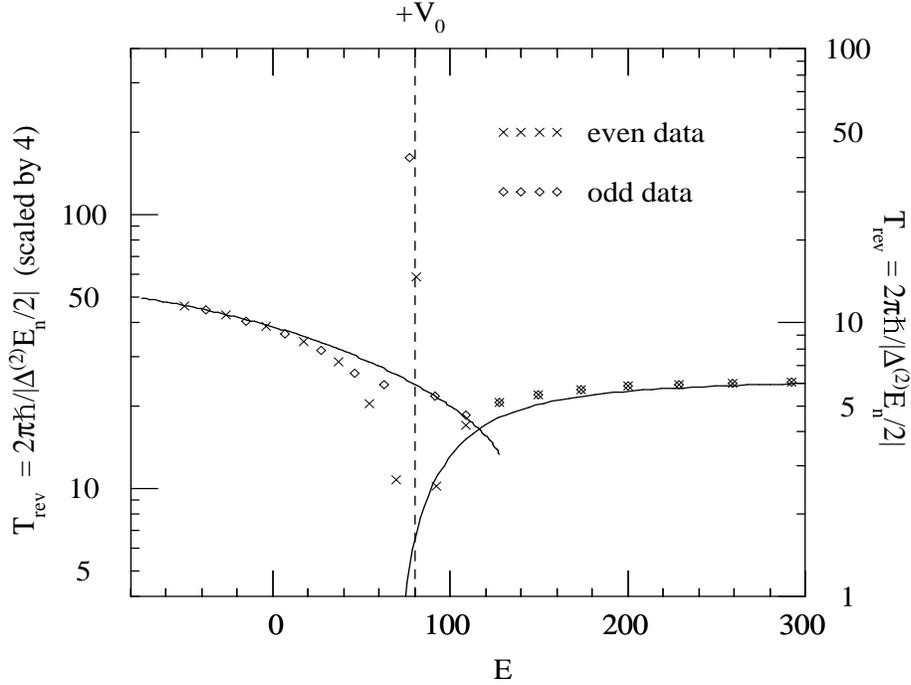,width=9cm,angle=270}
\caption{Plot of the revival times 
$T_{rev} \equiv 2\pi \hbar/|\Delta^{(2)} E_n/2|$ versus
$E_n$ for the parameters in Eqn.~(\ref{parameters}) for even (crosses)
and odd (diamonds) energy eigenvalues. The solid curves are the
predictions using 4th-order perturbation theory from the respective
limiting cases.}
\end{figure}

\end{document}